# Origin of Metallic States at Heterointerface between Band Insulators LaAlO$_3$ and SrTiO$_3$

K. Yoshimatsu[1], R. Yasuhara[1], H. Kumigashira[1,2,*], and M. Oshima[1,2]

[1]*Department of Applied Chemistry, University of Tokyo, 7-3-1 Hongo, Bunkyo-ku, Tokyo 113-8656, Japan*

[2]*Core Research for Evolutional Science and Technology (CREST), Japan Science Technology Agency, Tokyo 113-8656, Japan*

* Author to whom correspondence should be addressed; Electronic mail : kumigashira@sr.t.u-tokyo.ac.jp

**Abstract**

We have studied the electronic structure at the heterointerface between the band insulators $LaAlO_3$ and $SrTiO_3$ using *in situ* photoemission spectroscopy. Our experimental results clearly reveal the formation of a notched structure on the $SrTiO_3$ side due to band bending at the metallic $LaAlO_3/TiO_2$-$SrTiO_3$ interface. The structure, however, is absent at the insulating $LaAlO_3/SrO$-$SrTiO_3$ interface. The present results indicate that the metallic states originate not from the charge transfer through the interface on a short-range scale but from the accumulation of carriers on a long-range scale.

Heterointerfaces based on perovskite oxides have heralded the possibility of creating new multifunctional properties in ways that would not have been possible by using single-phase bulk materials [1–3]. Perovskite oxides with the chemical formula $ABO_3$ consist of AO and $BO_2$ atomic layers alternatively stacked along the *c*-axis. Thus, the novel physical properties that emerge at the heterointerfaces may depend on the atomic layer stacking sequence and the resultant charge redistribution near the heterointerfaces [4–6]. A particularly interesting aspect of these heterostructures seems to be the appearance of metallic conductivity with extremely high mobility at the interface between the band insulators $LaAlO_3$ (LAO) and $SrTiO_3$ (STO), depending on the terminating layer of the interface [7]. However, despite intensive theoretical and experimental studies [8–13], the mechanism of metallic-state formation is still unclear, and a large number of discussions exist on the fundamental origin of the high electrical conductivity and mobility. One of the scenarios proposed for describing the physics of the metallic states is the "charge transfer" that originates from the charge discontinuity at the interface between polar LAO and non-polar STO [7–9]. According to this scenario, in order to prevent the potential divergence in LAO films, electrons are transferred from the LAO layer into the $TiO_2$ bonds of the STO layer through the interface. Consequently, the "electron-doped" STO layer in the vicinity of the heterointerface begins to conduct. On the other hand, some recent experimental results have suggested that the "oxygen vacancy" in STO layers generated by the deposition of LAO films plays an important role in metallic-state formation [11,14,15]. However, if the LAO is grown on a SrO-terminated STO and not on $TiO_2$, the interface behaves as an insulator; this strongly suggests that the details of the interface do matter [16].

Photoemission spectroscopy (PES) is a powerful experimental technique for studying this issue, since it provides direct information on the electronic structure at heterointerfaces. Although metallic conductivity appears in the STO in both the abovementioned scenarios, the in-depth distribution of extra electrons on the STO side is expected to be significantly different. In the former case, electronic charges that are transferred partially from the LaO layer are confined in a few STO layers at the interface on a short-range scale. Consequently, partially filled Ti-$3d$ derived density of states (DOS) are expected to appear at the Fermi level ($E_F$) with a detectable photoemission signal owing to the interfacial sensitivity of PES measurements. On the other hand, since the electrons widely spread over STO layers in the latter case, the Ti $3d$ DOS should be negligibly weak (below the detection limit of the present PES measurement). In this Letter, we report the interfacial electronic structure of the LAO/STO multilayers, especially DOS near $E_F$ and the band discontinuity at the interface. We show that the metallic states originate not from the charge transfer but from the accumulation of carriers on a notched structure formed at the STO in the interfacial region. The interfacial band structure, as determined by the experiments, provide a comprehensive understanding the appearance of metallic conductivity at the interface, depending on the terminating layer of the interface.

LAO/STO multilayers with an $AlO_2$-LaO-$TiO_2$ or $AlO_2$-SrO-$TiO_2$ interface were fabricated on atomically flat $TiO_2$-terminated Nb:STO (001) substrates in a laser molecular-beam epitaxy chamber connected to a synchrotron radiation photoemission system at BL2C of the Photon Factory, KEK.  Nb doping of the substrates was necessary for preventing the charging effects in the PES measurements.  During deposition, the substrate temperature was kept at 700°C and the oxygen pressure was $10^{-5}$ Torr.  The LAO layer thickness was varied from 0 to 6 ML, while the thicknesses of the "buffer" STO layers were fixed at 20 ML to inhibit the influence of substrates. For the metallic LAO/$TiO_2$-STO interface, we directly deposited LAO layers on the STO layer, while LAO/SrO-STO was fabricated by depositing a SrO monolayer between the LAO and STO depositions.  During growth, the thickness of each layer was controlled on an atomic scale by monitoring the intensity oscillation of the specular spot in reflection high-energy electron diffraction.  These samples were subsequently annealed at 400°C for one hour and under atmospheric pressure of oxygen to fill residual oxygen vacancies.  After deposition, the samples were transferred into a photoemission chamber under ultrahigh vacuum of $10^{-10}$ Torr.  The PES spectra were recorded using a Scienta SES-100 electron energy analyzer with a total energy resolution of 150 – 200 meV in the 600 – 800 eV photon energy range.

Atomic force microscopy and electrical resistivity measurements were performed for characterization.  Atomically flat step-and-terrace structures were clearly observed in all the measured multilayers.  This indicated that the intended interface was controlled on an atomic scale. We also confirmed the formation of a chemically abrupt interface between the LAO and STO layers by comparing the relative intensities of the core levels as a function of the LAO cover layer thickness using a simulated photoelectron attenuation function [17]. For electrical resistivity measurements, we confirmed the metallic conductivity of LAO/STO interfaces fabricated on $TiO_2$-terminated STO substrates under the same growth conditions as for the PES measurements. The results are shown in the inset of Fig. 1 (b). In comparison with the previous studies [7, 11, 14, 15], the sheet charier density in the present samples is evaluated to be the order of $10^{13}$ - $10^{14}$ cm$^{-2}$.

Figure 1 shows the valence band spectra of metallic LAO/$TiO_2$-STO multilayers where the top layer was that of LAO with variable thickness. The valence band spectra mainly consist of two prominent O-2$p$ derived structures located in the binding energy range of 3.0 – 9.0 eV. Each PES spectrum is normalized to the integrated intensity of the O-2$p$ derived structures. The valence band spectra show systematic changes with an increase in the LAO thickness. The additional structures within the band gap, which are commonly observed at oxide surfaces with defects and/or disorders [18,19], are hardly visible in the present spectra, indicating that a well-defined surface (interface) is preserved through fabrication and in-vacuum sample transferring processes. The valence band maximum (VBM) of STO is located at 2.9 – 3.0 eV, which is defined as the energy position of the intersection between the extrapolated slope from the O 2$p$ states and the background. Since the band gap of STO is 3.2 eV at room temperature [18], the conduction band minimum (CBM) is located at 0.2 – 0.3 eV above $E_F$.

In contrast to the systematic changes in O-2$p$ derived states in the valence band spectra, there are no detectable changes near $E_F$ even with increasing LAO thickness. It is well established that Ti 3$d$ derived DOS emerges at $E_F$ when electrons are doped into STO [20,21]. Since LAO is an insulator with a considerably larger band gap (5.6 eV) as compared to STO (3.2 eV), LAO overlayers can be regarded as "transparent" for the electronic structure near $E_F$ of buried STO layers. Thus, we directly address the evolution of the metallic states at the interface [22]. If it is assumed that there is a reduction in the Ti states due to the charge transfer of an extra half electron per two-dimensional unit cell from the charge donor LaO layer to the adjacent $TiO_2$ layer, significant Ti-3$d$ derived states should appear near $E_F$. However, we did not observe any indication of Ti 3$d$ states within the experimental error. The absence of Ti-3$d$ derived states is further confirmed by Ti 2$p$→3$d$ resonant photoemission measurements (indicated by thick lines in Fig. 1(b)), wherein the photoionization cross section for Ti 3$d$ states is strongly enhanced [22]. In addition, the chemical stability of the $Ti^{4+}$ states in the topmost $TiO_2$ layer, irrespective of the neighboring donor layer, was also confirmed by Ti 2$p$ core level PES measurements.

The absence of Ti 3$d$ states near $E_F$ is clearly evident from a comparison of the PES spectra near $E_F$ with numerical simulations based on the charge transfer scenario [7,8]. In the simulations, we assumed that (i) half an electron per two-dimensional unit cell moves from the LaO layer just at the interface to the STO layers, and the extra electrons are accommodated in Ti 3$d$ bands; (ii) the doped electrons are confined in the adjacent TiO$_2$ layer; and (iii) the valence bands of LAO and non-doped STO consist of O 2$p$ states due to their ionic bond character.  In addition, the semiellipsoid-shaped DOS are used for roughly representing the Ti 3$d$ bands, which were observed in a previous PES study on La$_{1-x}$Sr$_x$TiO$_3$ [21].  The hypothetical Ti 3$d$ DOS are calculated by taking into account the photoelectron attenuation functions for LAO and STO layers [23] and the ratio of photoionization cross section between O 2$p$ and Ti 3$d$ states [24].  By comparing the numerical simulations with the experimental results, we concluded that the electron confinement proposed by the charge transfer scenario does not occur at the LAO/STO heterointerface.

The present results lead to two possibilities regarding in-depth carrier distribution.  The first is that the decay length of electrons generated by the charge transfer is too long to be detected by the present PES measurement.  The second is that the carrier density responsible for metallic states is too low.  For discussing the origin of the metallic states in the LAO/STO heterointerface, it would be worthwhile to roughly estimate the decay length of the generated carrier as well as the carrier density that can be probed by the present PES experiment.  In Fig. 2(b), we present the results of the simulation of Ti 3$d$ DOS where the exponential decay length, $\lambda$, was varied from 0.5 nm to 5 nm.  Since the metallic conductivity commonly appears above the critical LAO thickness [10], we chose the 4-ML spectra for comparison. The detection limit is indicated by a bar in the figure, and it is determined by the standard deviation of the noise signal from a constant background. Judging from the comparison between the experimental data and simulations, the decay length of the carriers is estimated to be greater than 5 nm. We also estimated the carrier density in the STO layers in the same manner and found that the carrier density of the interface accessible for PES measurements (~1 nm away from interface) is less than $10^{13}$ cm$^{-2}$.

The recent theoretical calculation based on the density functional theory revealed that the extra electrons from the donor LaO layer are redistributed in STO with a decay length of about 1 nm [24], leading to the formation of two-dimensional electron confinement at the LAO/STO heterointerface. In fact, the decay length of 1 nm has been observed in scanning transmission electron microscopic studies on the charge redistribution of STO/LaTiO$_3$ superlattices [4]. The predicted decay length of 1 nm is much shorter than the present results. Although the calculated decay length is modulated by the electron-phonon interaction with ionic polarization, the much longer decay length estimated by the present study strongly suggests that the charge transfer from LAO to STO does not occur on a short-range scale. The carriers responsible for the metallic states in LAO/STO may widely spread over the STO layers.

Next, we will discuss the origin of the metallic states in the LAO/STO heterointerface in terms of the discontinuity in the band structure between the two insulators.   To address the band discontinuity, we measured the shift of the Ti 2*p* core level as a function of the LAO overlayer thickness for the metallic LAO/TiO$_2$-STO interface as well as the insulating LAO/SrO-STO interface.   We did this because the band bending induced by the deposition of LAO on STO can be directly determined from the core level shifts of each layer.   The results are shown in Fig. 3.  Interestingly, the metallic LAO/TiO$_2$-STO interface clearly shows a peak shift toward higher binding energy as the LAO overlayer thickness increases, and the shift nearly saturates above 4 ML [25].   In contrast, the insulating interface does not show any detectable shift.   Judging from the saturation level of the core level shifts, the energy shift due to band bending for the metallic interface can be estimated to be -0.25 ± 0.07 eV.   Since the CBM in STO is located at 0.2 – 0.3 eV above $E_F$, as mentioned previously, the results strongly suggest that the CBM in STO at the metallic interface is nearly attained at $E_F$.

A band diagram of the metallic LAO/STO heterointerface, as deduced from the PES experiments, is illustrated in Fig. 4.   The existence of the notched structure inside the STO layers suggests a mechanism that produces the metallic states while still avoiding the potential divergence in the polar LAO layers.   Since the polar LAO layers have alternating ±*e* charge sheets, where *e* is the electron charge, the stacking of the LAO layer on non-polar STO produces a positive electric field.   This in turn leads to an electric potential that diverges with an increase in the LAO overlayer thickness.  The divergence catastrophe can be avoided by the formation of the long-range electric potential inside the STO whose spatial variation is governed by the carriers in the STO layers.  In other words, the accumulation of electrons in the notched structure, where the electrons may be generated by oxygen vacancies in the STO layers [11,14,15], produces the metallic states at the LAO/STO heterointerface.  This situation serves as a good analogy to the conductive channel of the ferroelectric field effect transistor after voltage application [26].  Although the origin of the band offset at metallic LAO/STO interfaces is not clear presently, the band structure determined in this work reasonably explains the characteristic features observed at LAO/STO interfaces: the appearance of metallic states above the critical thickness [9,10] and under an applied gate voltage [10], the insulating behavior observed by inserting an SrO atomic layer between two insulators, and the high carrier mobility which is similar to that of STO [7].

In conclusion, we have performed *in situ* PES study on LAO/STO multilayers to investigate the interfacial electronic structure responsible for the anomalous metallic states in the LAO/STO interface.  We clearly found that the notched structure was formed in the STO layers in the interfacial region, depending on the terminating layer of the interface.  These results indicate that the metallic conductivity originate not from the charge transfer through the interface but from the accumulation of carriers on the notched structure at the interface.

We are very grateful to M. Lippmaa for useful discussions.  This work was supported by a Grant-in-Aid for Scientific Research (A19684010 and A19204037) from JSPS.

**Figure captions**

**FIG. 1.** (color online)  (a) Valence-band photoemission spectra of metallic LaAlO$_3$/TiO$_2$-SrTiO$_3$ multilayers obtained by varying the LaAlO$_3$ overlayer thickness. (b) The spectra near $E_F$ measured with a smaller energy interval and a higher signal-to-noise ratio (dots).  The black triangle indicates the valence band maximum of SrTiO$_3$ layers.  The Ti-3$d$ state sensitive Ti 2$p$-3$d$ resonant photoemission spectra (recorded at photon energy of 466 eV) are superimposed as thick lines.  The inset shows the temperature dependence of electrical resistivity for the metallic LAO (15 ML)/ TiO$_2$-STO interface.  Clear metallic behavior is observed, indicating the existence of metallic conductivity in the present LAO/STO heterointerfaces.  Note that we could not measure the resistivity of insulating LaAlO$_3$/SrO-SrTiO$_3$ interfaces owing to the limitation of our apparatus.

**FIG. 2.** (color online)  (a) Comparison of PES spectra near $E_F$ (dots) with the numerical simulations based on the charge transfer scenario (thick lines).  (b) Comparison of PES spectra with the simulation for the 4-ML LaAlO$_3$ overlayer, where the transferred half electron is distributed from the interface with an exponential decay length of 0.5 – 5 nm, as shown in the inset.  The bar shows the detection limit of the present PES experiments.

**FIG. 3.** (color online)  Band bending of SrTiO$_3$ layers for metallic and insulating

LaAlO$_3$/SrTiO$_3$ interfaces: (a) Ti 2p core level spectra of a TiO$_2$-terminated SrTiO$_3$ and a metallic LaAlO$_3$/TiO$_2$-SrTiO$_3$ interface. (b) Spectra of a SrO-terminated SrTiO$_3$ and an insulating LaAlO$_3$/SrO-SrTiO$_3$ interface. (c) Plots of the energy shift of the Ti 2p core-level peaks for metallic (red) and insulating (blue) LaAlO$_3$/SrTiO$_3$ interfaces as a function of the LaAlO$_3$ overlayer thickness.

**FIG. 4.** (color online) Band diagram of the metallic LaAlO$_3$/TiO$_2$-SrTiO$_3$ interface determined by the present experiments. The band gaps of SrTiO$_3$ and LaAlO$_3$ are 3.2 eV and 5.6 eV, respectively. The valence band maximum (VBM) of SrTiO$_3$ is located at 2.9 – 3.0 eV. The VBM is nearly continuous between SrTiO$_3$ and LaAlO$_3$, which is confirmed by the almost constant energy position of the leading edge of the valence band structures (see Fig. 1(a)). As a consequence of downward band bending, a notched structure is formed in the SrTiO$_3$ layer in the interfacial region.

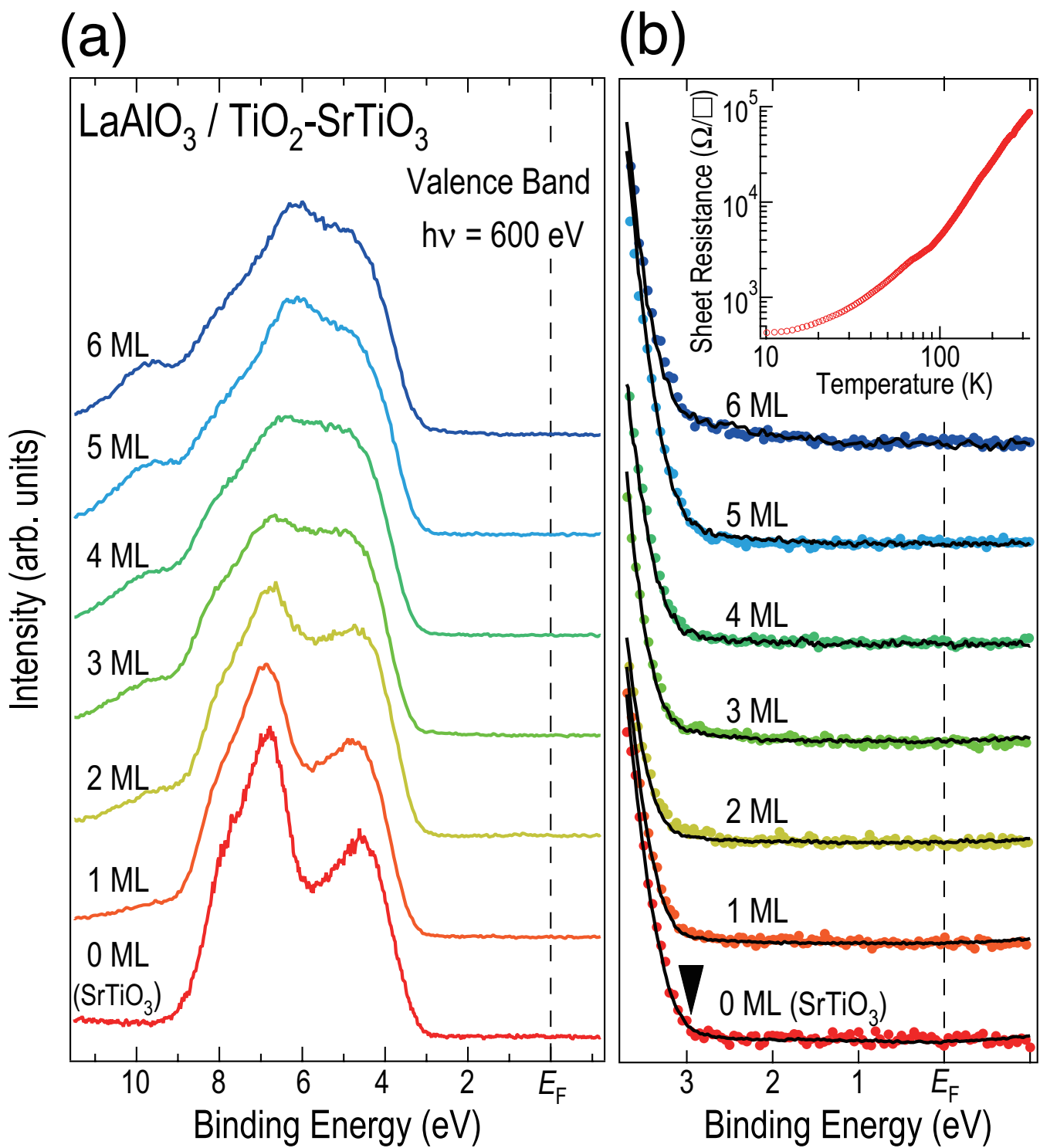

Figure 1

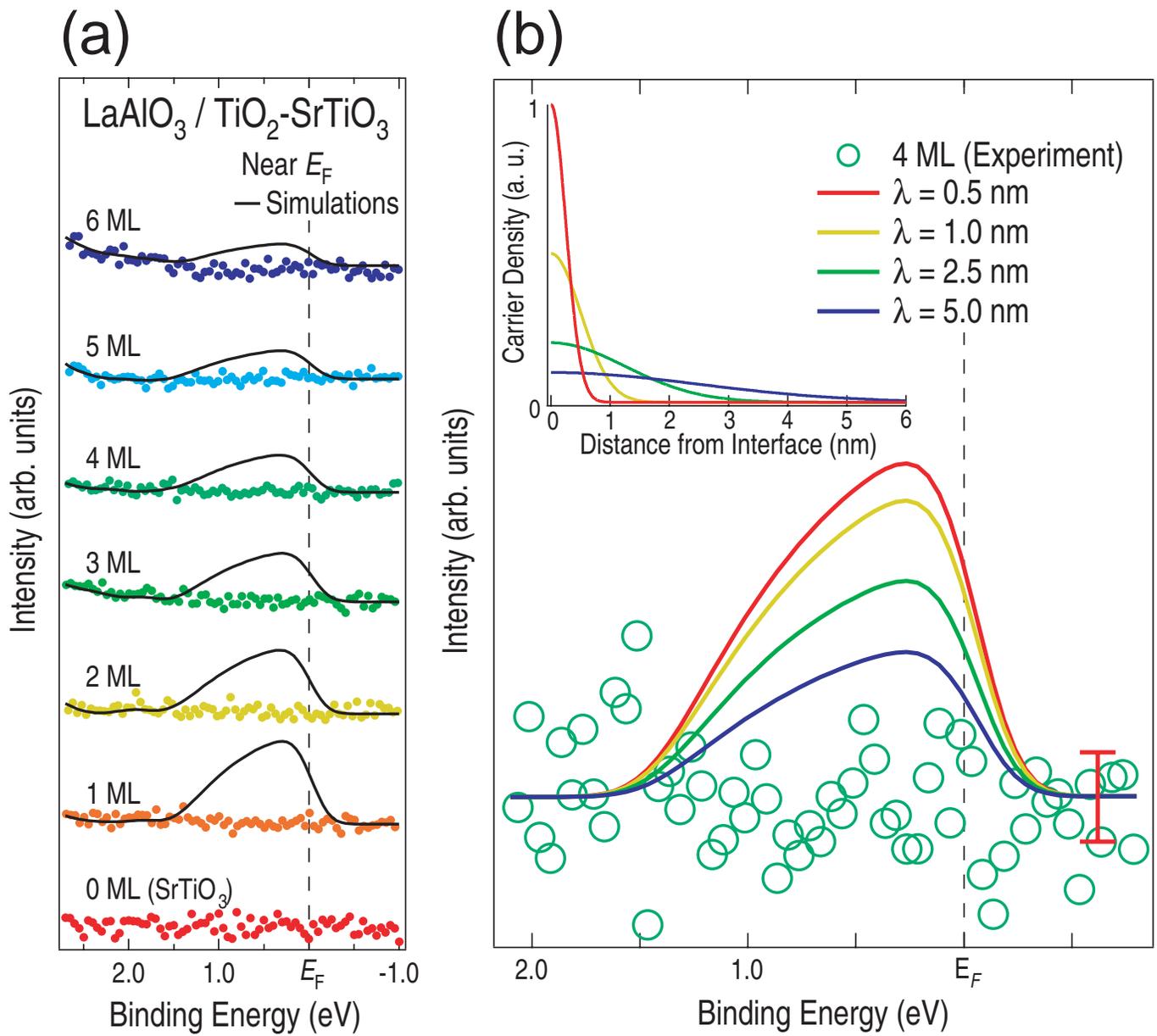

Figure 2

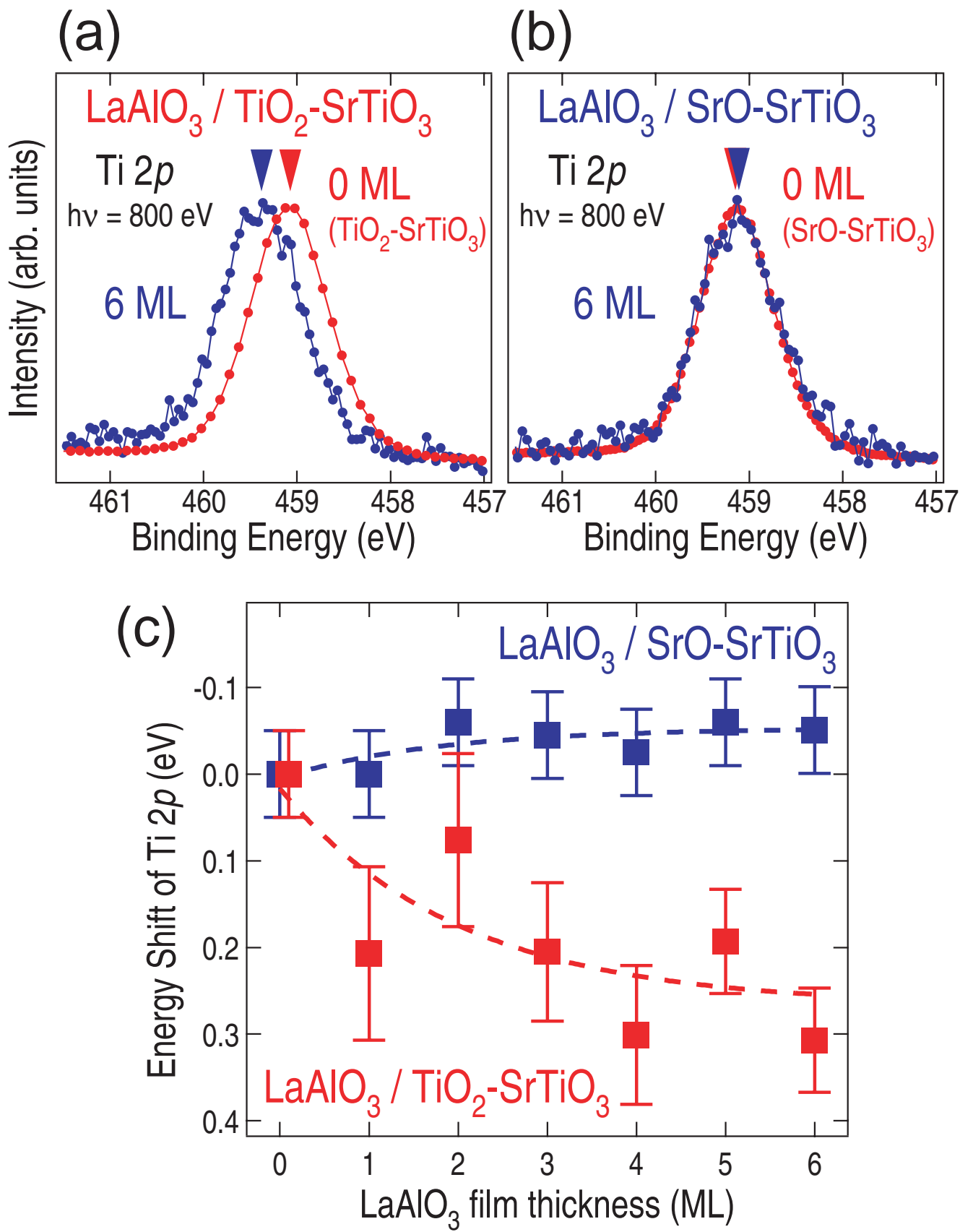

Figure 3

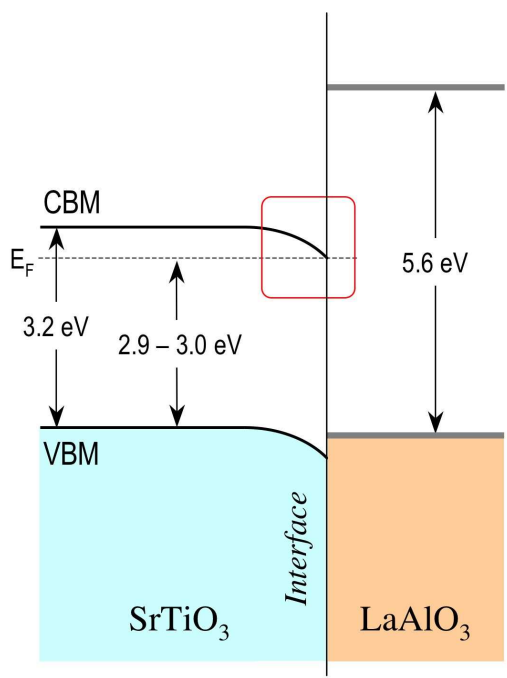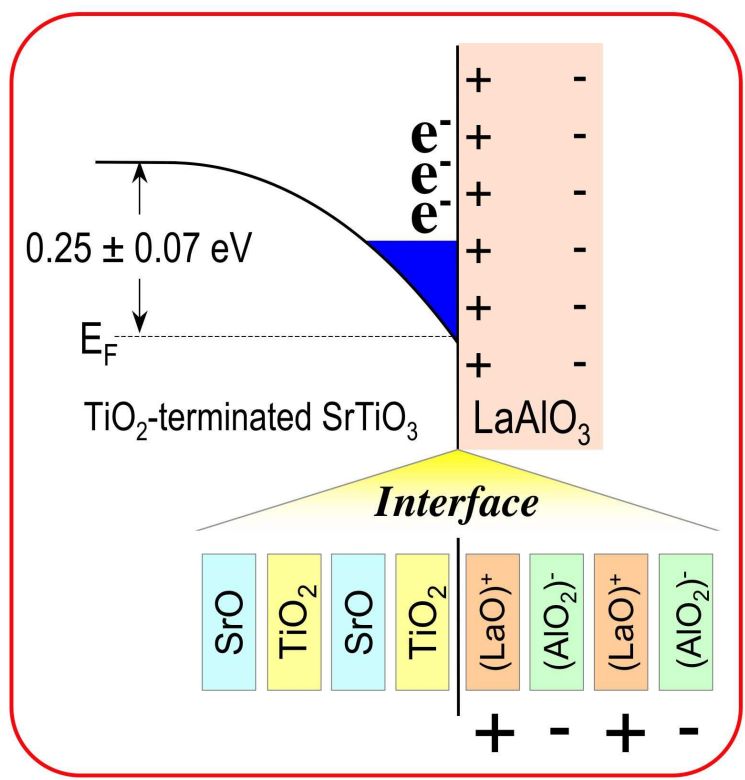

Figure 4